\documentclass{sigchi-ext}
\usepackage[T1]{fontenc}
\usepackage{textcomp}
\usepackage[scaled=.92]{helvet} 
\usepackage{graphicx} 
\usepackage{balance}  
\usepackage{booktabs} 
\usepackage{ccicons}  
\usepackage{ragged2e} 

\usepackage{array}
\newcolumntype{L}[1]{>{\raggedright\let\newline\\\arraybackslash\hspace{0pt}}m{#1}}
\newcommand{\squeezeup}{\vspace{-0.5cm}}

\usepackage{balance}
\usepackage{cite}
\usepackage[all]{nowidow}
\usepackage{hyperref}
\usepackage[dvipsnames]{xcolor}
\def\plaintitle{Revealing Neural Network Bias to Non-Experts Through Interactive Counterfactual Examples} 
\def\emptyauthor{}
\def\plainkeywords{neural networks, bias, interactive visualization}

\title{Revealing Neural Network Bias to Non-Experts Through Interactive Counterfactual Examples}

\numberofauthors{6}
\author{%
  \alignauthor{%
    \textbf{Chelsea M. Myers}\\
    \affaddr{Drexel University} \\
    \affaddr{Philadelphia, PA, USA} \\
    \email{chel.myers@gmail.com} }\alignauthor{%
    \textbf{Anushay Furqan}\\
    \affaddr{Drexel University} \\
    \affaddr{Philadelphia, PA, USA} \\
    \email{anushay.furqan@gmail.com} } \vfil \alignauthor{%
    \textbf{Evan Freed}\\
    \affaddr{Drexel University} \\
    \affaddr{Philadelphia, PA, USA} \\
    \email{efreed52@yahoo.com} }\alignauthor{%
    \textbf{Sebastian Risi}\\
    \affaddr{IT University of Copenhagen} \\
    \affaddr{Copenhagen, Denmark} \\
    \email{sebr@itu.dk} } \vfil \alignauthor{%
    \textbf{Luis Fernando Laris Pardo}\\   
    \affaddr{IT University of Copenhagen} \\
    \affaddr{Copenhagen, Denmark} \\
    \email{lula@itu.dk}}\alignauthor{%
    \textbf{Jichen Zhu}\\
    \affaddr{Drexel University} \\
    \affaddr{Philadelphia, PA, USA} \\
    \email{jichen.zhu@gmail.com} } }

\definecolor{linkColor}{RGB}{6,125,233}
\hypersetup{%
  pdftitle={\plaintitle},
  pdfauthor={\emptyauthor},
  pdfkeywords={\plainkeywords},
  bookmarksnumbered,
  pdfstartview={FitH},
  colorlinks,
  citecolor=black,
  filecolor=black,
  linkcolor=black,
  urlcolor=linkColor,
  breaklinks=true,
}

\begin{document}


\CopyrightYear{2020}
\setcopyright{rightsretained}
\copyrightinfo{}

\maketitle

\squeezeup

\RaggedRight{} 

\begin{abstract}
AI algorithms are not immune to biases. Traditionally, non-experts have little control in uncovering potential social bias (e.g., gender bias) in the algorithms that may impact their lives. We present a preliminary design for an interactive visualization tool, \textit{CEB}, to reveal biases in a commonly used AI method, Neural Networks (NN). \textit{CEB} combines counterfactual examples and abstraction of an NN decision process to empower non-experts to detect bias. This paper presents the design of CEB and initial findings of an expert panel  ($n=6$) with AI, HCI, and Social science experts. 
\end{abstract}





\printccsdesc

\section{Introduction}

Artificial Intelligence (AI) methods are used in increasing range of aspects of our daily lives. As the adoption of these AI applications widens, the need for transparency and accountability becomes more pressing. Governments have started to require companies to be transparent about AI applications with social significance. For example, {\em profiling models}, a widely used method to model certain aspects of a person ~\cite{Goodman2016} (e.g., financial creditworthiness), raised public concerns.  
Reports show biased profiling models can produce devastating consequences for those it unfairly models, when for example deciding the risk of recidivism for parole~\cite{Dressel2018, Tan2017} or which patients receives extra care ~\cite{Johnson2019RacialPatients}.

Currently, the general public relies on AI experts to discover these biases. However, this approach cannot easily scale up with the rapid adoption rate of AI applications. More important, technical experts may not be fully aware of the needs of the communities about which AI algorithms are making decisions. As a result, there is a recent call to empower non-experts to open the blackbox of AI and better understand its decision making process ~\cite{Amershi2014, Yang2018ux, Xie2019}. 

\begin{margintable}[1pc]
  \begin{minipage}{\marginparwidth}
    \centering
    \small
    \begin{tabular}{L{4cm}}
      {\small \textbf{Features}} \\
      \toprule
        Gender \\ \hline 
        Education \\ \hline 	
        Self-Employment \\ \hline 
        Income \\ \hline 
        Credit History \\ \hline 
        Requested Loan Amount \\ \hline 	
        Requested Loan Duration	\\  
      \bottomrule
    \end{tabular}
    \caption{The application features in our dataset and used in \textit{CEB's} NN.}~\label{tab:feat}
  \end{minipage}
\end{margintable}

In this paper, we present an interactive visualization for AI non-experts to explore a semantic Neural Network's (NN) decisions, in the context of profiling models for loan applications, to reveal potential bias. Among AI methods, NN-based ones are notorious for its low interpretability \cite{Zhu2018}. 
Our tool, \textit{C}ounterfactual \textit{E}xamples for \textit{B}ias (\textit{CEB}), is designed for non-experts to discover potential biases. It explores the approach of 1) visualizing activation patterns of the NN to increase its interpretability, 2) using counterfactual example to facilitate non-experts to discover biases in the algorithm. By visualizing how counterfactual examples may impact the decision made by an NN, \textit{CEB} aims to facilitate non-experts to decide if bias is present. 
To our knowledge, this work is among the first tools that support non-experts to find bias in NN algorithms.


Employing an iterative and human-centered approach, we have built a prototype of \textit{CEB} and reviewed its design through interviews with AI, HCI, UX, and Sociology experts. 
In the rest of the paper, we present \textit{CEB} and the results of our expert panel with six experts. Overall, we found the experts believed \textit{CEB} would be an intuitive tool for non-experts. Experts' believed the counterfactual examples highlighted bias while the abstraction of datapoints into clusters allowed users not to be overwhelmed by the sample size.

\section{Related Work}


Bias occurs in a variety of domains such as emotion recognition~\cite{Howard2017}, word embeddings~\cite{Garg2018WordStereotypes,Bolukbasi2016DebiasingEmbedding}, and object classification~\cite{Zhao2017MenConstraints}. Research addressing algorithmic bias typically alters or supplements algorithms to correct bias~\cite{Howard2017, Kim2018LearningData, Alvi2019TurningEmbeddings, Das2019MitigatingApproach, Amini2019UncoveringStructure, Kusner2017CounterfactualFairness}. These automated approaches can reduce the unfairness of algorithms; however, they can trade-off accuracy and still do not guarantee complete fairness~\cite{Kusner2017CounterfactualFairness}. 

We argue that supporting users to detect bias, instead of algorithms, is an alternate approach when automation is not available or feasible. This approach supports the General Data Protection Regulation (GDPR), highlighting people's right to algorithmic explanations and non-discriminating algorithms\footnote{See the GDPR Articles 13-15 and 22 for more information.}. Research on Interactive Machine Learning (IML) and eXplainable AI (XAI) often designs for algorithmic explanations as well. When designing \textit{CEB,} we first looked towards these fields' findings.
Research on IMLs has developeded techniques helpful to both experts and non-experts in understanding a Machine Learning (ML) model. Interacting with a ML model in an IML can assist non-experts in learning data requirements of a model and develop more realistic expectations of its capabilities~\cite{Fiebrink2011}. Research on AI education for non-experts focuses on increasing their understanding of how specific models work to empower their use as a design material~\cite{Dove2017, Yang2018, Yang2017}. We argue that educating non-experts on the social implications of AI, specifically bias, is another pressing issue that is currently under-researched. Projects in XAI develop various UI explanations to aid experts, and occasionally non-experts, in understanding AI decisions~\cite{Abdul2018,Adadi2018, Zhu2018}. UI techniques such as natural language explanations~\cite{Ribeiro2016} and comparative and normative examples for image classification~\cite{Cai2019} have been found to help non-experts understand a NN's decision. Similar to IML, interactive explanations have been found to increase non-expert's objective and self-reported understanding of the profiling model but require more of a user's time~\cite{Cheng2019}. Tools similar to \textit{CEB} aim to explain ML models to children~\cite{Hitron2019} or game designers~\cite{Xie2019}, but do not focus on potential bias.

We emphasize abstraction and counterfactual examples to facilitate the discovery of NN bias. Abstracting the ML process has been found to assist the understanding of non-experts~\cite{Yang2018ux,Dove2017}. Our abstraction is based on reducing and plotting the hidden node activations of a NN; a technique used in tools to reveal the ``black-box'' of image classification NNs~\cite{Carter2019, Olah2018}. We further abstract these activations by clustering them, a technique commonly used in data visualization to improve interpretability~\cite{Ma2019ExplainingAnalytics, Ma2018ScatterNet:Scatterplots, Liao2018Cluster-BasedScatterplots}.
Employing counterfactual examples is a technique seen in developing more fair models~\cite{Kusner2017CounterfactualFairness, Wachter2017, sokol2019counterfactual}. \textit{CEB} focuses on illustrating the potential bias of a NN through counterfactual examples since they have been shown to improve a non-experts understanding of AI concepts~\cite{Ribera2019, Wachter2017}.


\begin{marginfigure}[-15pc]
  \begin{minipage}{\marginparwidth}
    \centering
    
    \vspace{6cm}
    \frame{\includegraphics[width=0.9\marginparwidth]{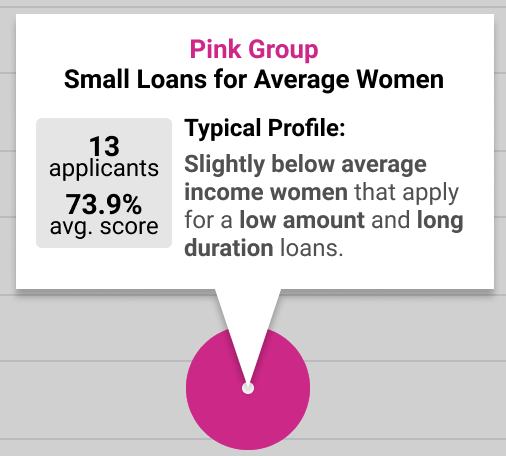}}
    \caption{Example of the natural language description and score}~\label{fig:nl}
    
    \frame{\includegraphics[width=0.9\marginparwidth]{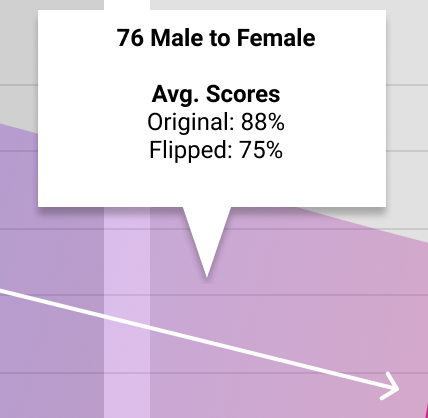}}
    \caption{Example of path score from original \textbf{{\color{Plum}Purple Group}} to flipped \textbf{{\color{WildStrawberry}Pink Group}}.}~\label{fig:path}
    
    \frame{\includegraphics[width=0.9\marginparwidth]{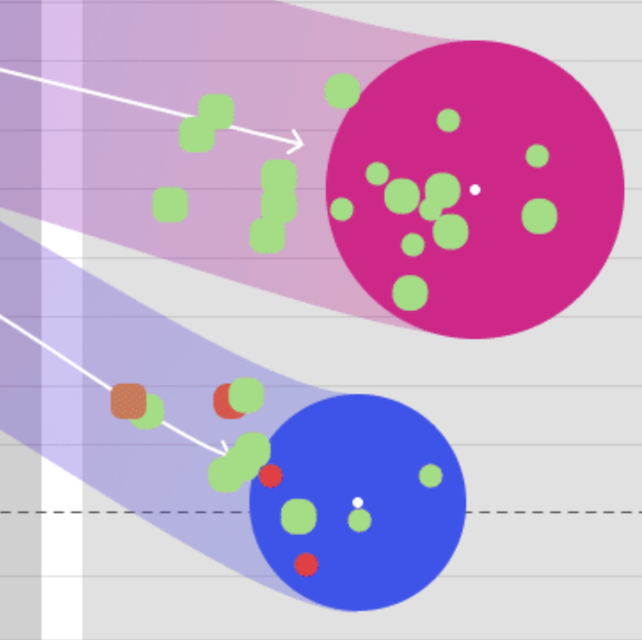}}
    \caption{Still frame of animated datapoints switching from original to flipped groups.}~\label{fig:anim}

  \end{minipage}
\end{marginfigure}

\section{Designing CEB to Facilitate Bias Detection}
We selected a pre-existing loan application dataset\footnote{Available at \href{https://www.kaggle.com/burak3ergun/loan-data-set}{https://www.kaggle.com/burak3ergun/loan-data-set}.} since this data already suffered from sampling bias (with a disproportionately higher amount of men represented than women). It contains datapoints about features of each load application (Tab.~\ref{tab:feat}) and its outcome (accept or reject). 
To prepare the training data, we first removed datapoints with missing information, reducing the dataset from 614 to 480. We then randomly divided the data into $2/3$rds for training and $1/3$rd for testing. The employed NN is a Fully Connected Neural Network (FCNN) with three hidden layers. The network has seven inputs (Tab.~\ref{tab:feat}) and one output neuron (a loan application is recommended for approval if the neuron’s output is higher than a threshold of 0.5). The activation function used for each layer of the network is the ReLu function except for the final output, which employs a sigmoid function. The NN was trained and modified until performance reached an accuracy of 79\%. This accuracy is competitive compared to other public models working with the same dataset\footnote{Examples of other model's accuracy can be found at \href{https://datahack.analyticsvidhya.com/contest/practice-problem-loan-prediction-iii/lb?page=1}{datahack.analyticsvidhya.com/contest/practice-problem-loan-prediction-iii/lb?page=1} and \href{https://www.kaggle.com/burak3ergun/loan-data-set/kernels}{www.kaggle.com/burak3ergun/loan-data-set/kernels}.}. 

\begin{figure*}
    \centering
    \vspace*{-3cm}
     
    \hspace*{-5cm}\includegraphics[width=25cm]{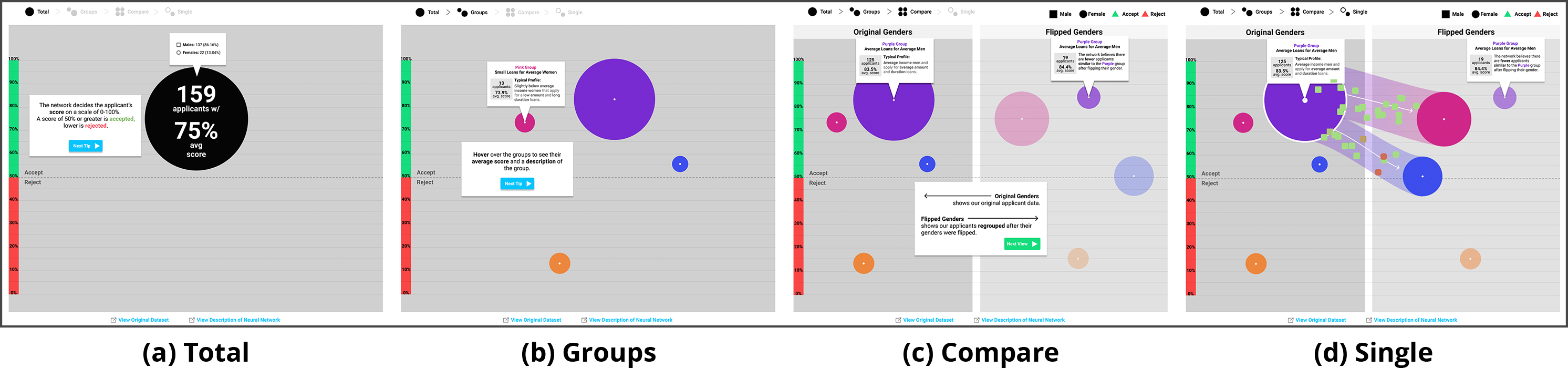}
    
    \hspace*{-5cm}\begin{minipage}{25cm}
      \caption{The four views of \textit{CEB}: 
    (a) \textit{Total} view providing an overview of the dataset, 
    (b) \textit{Groups} view showing the NN splitting the dataset into four clusters, 
    (c) \textit{Compare} view with counterfactual example by regrouping datapoints after flipping their gender feature,
    and (d) \textit{Single} view animating datapoints being regrouped into counterfactual example groups. 
    Interact with \textit{CEB} \href{https://www.figma.com/proto/Wtx4n2ZmlWjnYXCDWi1cBf/XAID}{here}.}\label{fig:views}
    \end{minipage}

    \squeezeup

\end{figure*}

In order not to overburden a non-expert user's cognitive load, \textit{CEB} abstract the dataset into clusters instead of directly visualizing individual datapoints (Fig.~\ref{fig:views}(b))~\footnote{In \textit{CEB}, we use the term ``group'' instead of ``cluster'' to avoid confusion by non-technical users.}. To do this, we first reduce the dimensionality of the NN activations to two dimensions with the T-distributed Stochastic Neighbor Embedding (t-SNE) \cite{VanDerMaaten2008VisualizingT-SNE} and then cluster the activations with k-means. With activations we mean the output that each neuron in the NN calculates, which is based on applying a non-linear function to the sum of the each neuron's  inputs. It is worth highlighting that we cluster based on the activations instead of the data itself, to gain insights into how similar or dissimilar the \textit{NN} interprets different groups of applicants. Since NN's activiation patters are correlated to the decision it makes, our technique of clustering activiations can reveal how the NN treats different applications. For instance, if women with high income are grouped with men with low income based on the NN's activation pattern, we can expect these two groups to have similar loan acceptance rates and hence investigate potential bias of the NN.

The NN's output is a score, on a scale of 0-100\%, to each application. A score of 50\% or higher means the loan application is accepted. Due to the low interpretability of NNs, we used counterfactual examples to help non-experts compare how similar applications may be treated by the algorithm. In particular, we explored the counterfactual examples created by changing one feature of an existing data point in the dataset. This allows users to ask questions such as ``If the same  application is from someone of a different race or gender, would the NN make the same decision about the loan?'' For our early prototype, we used binary gender as our focus. A user can compare clusters of applicants with their equivalent counterfactual examples where the genders of applicants are ``flipped.''



\textbf{\textit{Using CEB:}} Users are guided through four views of \textit{CEB}: \textit{Total}, \textit{Groups}, \textit{Compare}, and \textit{Single}. 
\textit{Total} (Fig.~\ref{fig:views}(a)) shows users a summary of all datapoints in the original dataset and their gender breakdown. We breakdown gender since this is the focus of the counterfactual example in this prototype of \textit{CEB}. 
\textit{Groups} (Fig.~\ref{fig:views}(b)) visualizes these datapoints splitting into clusters that the NN considers similar. 
Users can hover over the clusters to see a summary of their prototypical datapoints and the cluster's average score (Fig.~\ref{fig:nl}). The clusters' y-coordinates correspond with the average score of all applications in the cluster.
\textit{Compare} (Fig.~\ref{fig:views}(c)) presents the counterfactual example that flips the gender feature of all datapoints in a cluster. 
This side-by-side view shows the original dataset clusters (seen in the \textit{Group} view) and the flipped dataset clusters .
Users can compare the clusters' average scores and see if flipping the gender feature impacts said score or cluster size.
Finally, in \textit{Single} (Fig.~\ref{fig:views}(d)), users can click on an original cluster and see which cluster its datapoints move to, after the gender feature is flipped. To highight the movement, we animated the datapoints and show how they move from the original to the new clusters (Fig.~\ref{fig:anim}). 
Users can hover over the original and flipped clusters to read their descriptions, or see datapoint \textit{path scores} (Fig.~\ref{fig:path}) by hovering over the arrows to see how many datapoints moved, their genders, and average score.

\section{Expert Interview Methodology}
We invited experts of related domains to interact with a prototype of \textit{CEB} and to determine if the NN was biased. In the beginning of the interview, we gave experts the false information that there were two versions of the prototype, assigned at random: one that was biased and the other was not. This was done to avoid influencing the experts so that they can make their own judgement if there were bias in the NN or not. Each session was conducted separately and began with a pre-session survey gathering expertise, data literacy, and demographics. Experts were allowed to explore the tool for a maximum of 20 minutes with the think-aloud protocol. After experts were satisfied with their conclusion (if the NN was biased or not), they were directed to a post-session survey and semi-structured interview asking if they believed their version of \textit{CEB}'s NN was biased and to provide evidence. Experts were encouraged to go back to the tool to refer to their evidence when speaking about it. Each session was also recorded and transcribed. To analyze the data, researchers reviewed the transcripts and survey data.

\section{Findings}
A breakdown of our experts is shown in Tab.~\ref{tab:experts}. We included two AI/ML experts in order to verify the scientific accuracy of our tool. The remaining experts we interviewed belong to our target user group - non-AI experts (non-experts in short). Overall, \textit{CEB} was well received by the experts. Experts commented on how this tool would help users to get a quick intuition on if bias was present. \textit{``It is a good visualization... it helps create intuitions in your head. Now you actually want to test those intuitions, right?''} [E4] Experts also commented on wanting more tools embedded in the visualization to analyze what other features may influence the NN's score. All experts were able to identify bias through the counterfactual example. This identification was made easier by the abstraction of datapoints into clusters. However, how the datapoints were clustered, based on activations or features, confused some experts.

\begin{margintable}[1pc]
  \begin{minipage}{\marginparwidth}
    \centering
    \small
    \begin{tabular}{L{0.5cm}|L{2.5cm}}
      {\small \textbf{\#}}
      & {\small \textbf{Expertise}} \\
      \toprule
        \textbf{E1} & Professor of Sociology   \\ \hline 
        \textbf{E2} & UX Research Director  \\ \hline 
        \textbf{E3} & UX Research Director  \\ \hline 
        \textbf{E4} & AI/ML Research Scientist  \\ \hline 
        \textbf{E5} & Professor of AI/ML  \\ \hline 
        \textbf{E6} & HCI Research Scientist  \\  
      \bottomrule
    \end{tabular}
    \caption{Expert reference numbers, position, and expertise. }~\label{tab:experts}
  \end{minipage}
\end{margintable}

\textbf{\textit{Counterfactual Example:}} 
Identifying bias through the comparison of the original and flipped clusters was facilitated using the y-coordinates of the clusters. All experts were able to isolate the change in the clusters' scores and conclude the presence of bias. Experts commented that the design choice to see NN scores go up or down on this axis was intuitive and provided jumping-off points to build a hypothesis for further exploration. Experts who skipped through first views and quickly went to the \textit{Single} view reported a better mental model of the redistribution of the datapoints from the original to the flipped clusters [E3, E5, E6] than those who spent more time on the first views  [E1, E2, E4]. These latter experts were confused about whether the datapoints stayed in their original clusters with their feature flipped or the datapoints flipped and moved to different clusters. \textit{Seeing} the \textit{Single} view's animation of the original datapoints being redistributed into the counterfactual clusters assisted experts in this understanding. Experts who spent more time on the first views without the animation of datapoints were unclear on what differences the counterfactual example presented. The confusion was resolved for all experts in the \textit{Single} view.

After using the y-axis to hypothesize bias, experts would rely on the clusters' score from the NN as more concrete evidence of bias. Second to this, experts relied on the path scores (Fig.~\ref{fig:path}) to isolate the specific score changes for men and women in these clusters. Unfortunately, the path scores were not noticed by all experts immediately [E2] or ever [E6] since the scores only appeared when hovering over the arrows between clusters in the \textit{Single} view. For the experts who did find it, they heavily relied on the scores as evidence of bias as well. The counterfactual clusters showed the averaged score of reclustered datapoints. E4 commented that the path scores allowed users to see more specific scores of the datapoints being reclustered to isolate bias. For example, the path scores allow a user to see a group of men being flipped to women and then seeing their original score was higher as men than as women. 

\textit{\textbf{Abstraction:}}
Experts found abstraction individual datapoints to activation clusters necessary in exploring and comparing the number of datapoints. However, since clustering features is a common approach, some experts were confused about how the clusters were formed. Experts with an AI background [E4, E5] were more likely to identify the clusters were based on the NN's activations. Experts without this expertise took longer to identify how these clusters were formed. E1 desired more explanation on why these were the most prominent clusters and wanted more context on how they were made. 

This issue was exacerbated by the natural language descriptions highlighting the cluster's average datapoint (Fig.~\ref{fig:nl}). This natural language was an important handle for experts to refer to the clusters to compare them. However, since the descriptions refer to features, this strengthened the confusion on whether clustering was based on activations or features. It is unclear if non-experts without this exposure to automated clustering would experience this same confusion. E1 suggested to add more explanation as to why these features were selected to demonstrate their impact on the cluster formation, if any.

\textit{\textbf{Other Comments:}}
Experts did enjoy the design and UX of \textit{CEB} and felt non-AI-experts would find it engaging and not overwhelming. Experts felt building tools such as this were crucial and \textit{"highly necessary...to work on explainability of neural networks and also to make tools for understanding bias."} [E5] E5 commented that the UX felt like a guided exploration mimicking working with data. Experts provided several other comments on \textit{CEB} as well. E4 pointed out that \textit{CEB} does not inform users from where users the bias comes. For example, if the bias comes from the dataset or the NN's model. A majority of experts [E1, E4, E5, E6] requested control over what feature the counterfactual example presented to explore other biases.

\section{Discussion \& Conclusion}
Overall, we found experts believed our tool -- using abstraction and counterfactual examples -- was a feasible approach to assist non-experts in detecting biased algorithms. The next step in our iterative design process is to revise \textit{CEB}'s design and evaluate it with non-experts. To strengthen \textit{CEB}, we will clarify 1) how clusters are formed, and 2) how datapoints are flipped and redistributed. 
First, to clarifying clusters, \textit{CEB} can further emphasize that the clusters are based on what datapoints the NN ``sees'' as being similar (activation). Tools visualizing activations are typically in the image classification domain and leverage this visual component to convey similarity~\cite{Carter2019, Olah2018} (e.g., users can \textit{see} an image of a cat looking similar to a small dog). Our semantic domain does not have the same advantage of being inherently visual. \textit{CEB} can instead rely on metaphors of the NN ``believing'' datapoints are similar to help overcome this. The features can be presented as what the NN was trained with to come to these beliefs. Animation can be leveraged to demonstrate grouping datapoints based on this ``belief'' when introducing the clusters. 
Second, to clarify datapoint distribution, the animation seen in the \textit{Single} view can be used in the \textit{Compare} view to build in the counterfactual clusters and show how the clusters are formed.
Lastly, we believe that the clusters can include more visual information to help users understand the characteristics of the datapoints it entails. For example, similar to Ma et al.~\cite{Ma2019ExplainingAnalytics}, the clusters' design can embed pie or radar charts to show a high-level distribution of datapoints' values across the features. This approach could also help lessen the negative impact of the natural language descriptions.

In summary, we presented our approach for facilitating non-AI-experts to discover diases in NN through our tool {\em CEB}. It attempts to do so by counterfactual examples and abstraction through clustering NN's activations. Current limitations are that it only presents one biased dataset, one feature change, and a relatively small dataset. Future versions of \textit{CEB} will present both bias and non-biased datasets. 
Important future work also includes studying how to guide non-experts' exploration of large datasets with more features. 

\balance{} 

\bibliographystyle{SIGCHI-Reference-Format}
\bibliography{references}

\end{document}